# Equatorial African Lightning: Past, Present and Future

Rohit Chakraborty[a], Parth Sanjeev Menghal[b*]

[a] Department of Physics and Nanotechnology, SRMIST, Chennai, India

[b] Department of Earth and Climate Sciences, Indian Institute of Science Education and Research Tirupati, India

**\* Correspondence:**

Parth Sanjeev Menghal

Department of Earth and Climate Sciences,

Indian Institute of Science Education and Research Tirupati, India

Email: menghalparths@gmail.com

## Abstract

Lightning strikes are one of the leading causes of death among natural disasters in tropical regions. The Congo rainforests host the highest rates of lightning flashes in the world and the lightning properties in this region have a strong seasonality owing to the cross-equatorial movements of the Inter Tropical Convergence Zone (ITCZ) throughout the year. The Lightning Flash Rates (LFR) are found to peak during the Equinoctial months while lightning radiance assumes the strongest values during the Boreal Summer months. Across the Tropical Rainfall Measuring Mission (TRMM) duration, (1998-2015), annual LFR in the Congo rainforest increases steadily, however, the average and the peak lightning radiances decrease by ~1%/year and ~1.5%/year respectively which is counter-intuitive to the expected intensification in weather extremes in a warming climate. An in-depth analysis highlighted the influence of moisture convergence over the Congo Basin from the Atlantic as the primary contributor to lightning compared to the aerosol nucleation and thermodynamic instability effects during the boreal summers. Further, it has been proposed that in a gradually warming climate, the strength of the prevailing ITCZ over Sahel region has intensified thereby weakening the moisture ingress towards the Congo Basin, and this resulted in the observed reduction in lightning intensity in recent years. Consequently, the long-term projections of lightning properties from global model simulations revealed a prominent weakening in lightning (~50%) over the Congo Basin by the year 2100 assuming the Representative Concentration Pathways (RCP) 8.5 emission scenarios.

## 1. Introduction

The Tropical band of 30º N-S hosts 78% of all lightning flashes around the globe (Christian et al. 2003). Thus, within the tropics, Lightning strikes generating from thunderstorm events pose great damage to infrastructure and the lives of humans and animals (Wierzchowski et al. 2002; Holle et al. 1999) hence they have always been a topic of great scientific attention (Chakraborty et al. 2014; Jana et al. 2023; Maitri et al. 2019; Rakshit et al. 2018; Saha et al. 2022). A significant factor behind the hazard of a lightning strike is the uncertainty in the position and timing of these strikes, which is primarily due to the spontaneity in the buildup of opposite charges. Hence, the frequency of lightning strikes (lightning flash rates) and their average intensity (lightning radiances) need to be better understood for effective protection against risks posed by lightning. Such lightning properties often correspond to the average particle composition of the clouds and their interactions in a convectively unstable atmosphere.

Lightning strikes are a product of the processes that govern the charge separation in clouds. Various atmospheric variables affect the above processes differently. For instance, convection strongly influences cloud electrification (Takahashi T 1978; Williams et al. 1985) by catalysing the interactions between ice and graupel particles inside the mixed layer. Similarly, the role of aerosols has also been discussed thoroughly in many past research attempts (Twomey et al. 1984; Westcott, 1995; Stallins and Rose, 2008; Rosenfeld and Lensky, 1998; Shi et al. 2020). Other than these, some studies have also attributed the role of orography to the strong density of lightning strikes along the windward slopes of the mountain ranges like the Himalayas and the northern Andes (Aggrawal et al. 2023; Muñoz et al. 2016).

Coming to the climatic trends, frequency of days witnessing intense lightning has been increasing, due to greenhouse gas induced surface warming (Finney et al. 2020). Precipitation is often linked to lightning occurrences due to similar origins, and in cases such as thunderstorms and cyclones, their strengths often correlate (Carte and Kidder 1977; Soula and Chauzy 2001; Adechinan et al. 2014). In addition, some attempts have depicted a ~40% increase in thunderstorm frequencies as a result of climate change happening across Central Africa, which again has been attributed to the instability as well as moisture trends (Harel and Price, 2020). While many variables are routinely mentioned in literature for their associations with lightning, cloud electrification is found to be dependent on surface pressure, updraft speed, liquid water content and cloud temperature, which is not necessarily linear (Saunders et al. 1991; Baker et al. 1999). Siingh et al. (2013) tried to understand the varying effect of various atmospheric variables on lightning by dividing India into regions according to their geography. While variables such as surface temperature, CAPE, Cloud Cover, rainfall, Outgoing Longwave Radiation (OLR) and sunspots showed expected correlations, a single variable did not dominate across all regions. Hence, these results support a hypothesis proposed in many past attempts wherein the lightning properties depend highly on its regional geographic and meteorological conditions (Orville and Huffines 2001; Finke 1996; Soriano et al. 2005).

Research attempts on satellite-based lightning distribution and climatology provided by Many studies (Christian et al. 2003; Boccippio et al. 2000) have depicted a global mean lightning frequency of 44 flashes/m^2/year. The use of OTD in these studies also finds that the highest lightning flash rates occur in the tropics along the coasts, mountain slopes and synoptic convergence zones. Additionally, lightning flash rates over land are found to exceed that over the sea by almost a factor of 8. The Association between ITCZ and lightning flashes has long been studied due to a similar set of atmospheric properties associated with both phenomena, namely moisture and convection (Collier and Hughes 2011). Mackerras et al. (1998) exclaimed that lightning between 5º N-S has a distinct semi-annual component. This corresponds to a twice-a-year peak in lightning flashes with different dynamics for each (Levinson 2005). A similar association was discounted over oceans due to lesser convection over the ocean (Williams et al. 1992; Zipser and Lutz 1994). Over land, peak lightning activity is found to be located at the southern edge of the ITCZ , and this is also supported over the Congo basin as it receives several thunderstorms due to low-level convergence of ITCZ while stationed over this region (Soula and Chauzy 2001). Notably, during this time, dry air from North Africa converges with the moist air from the Atlantic, creating ideal conditions for intense convection and electrification (Collier and Hughes 2011).

Recent lightning studies using coarse resolution over Africa reveal that the eastern Congo Basin experiences the most lightning flashes globally, with an average of 160 flashes/km/yr (Cecil et al. 2014). Another candidate for the highest flash rates has been reported to be Maracaibo Lake in North Venezuela. Notably, both of these places harbour high lightning flash rates due to the local impact of wind dynamics and topography. However, the distribution of lightning across this region are modulated by a series of many factors, like the ITCZ movement and the influence of African Easterly jets (Albrecht et al. 2016), while that in Lake Maracaibo corresponds to a narrow valley along a tropical coast. Yet another study using WWLLN data in the Congo basin mentions maximal lightning activity in December and January and minimal activity in June, July and August (Soula et al. 2016). In view of this, several studies mention various pathways of moisture influx into the Congo basin, such as the bi-yearly passage of ITCZ over the Congo basin, which draws moisture from the Gulf of Gabon into the Congo basin. The Virunga mountains are also presumed to receive a fraction of their moisture supply from the Rift Valley lakes to the east of the mountains (Collier and Hughes 2011). Thus, the Congo basin boasts frequent MCSs (Mesoscale Systems) with much higher cloud liquid and ice content, thereby leading to greater electrification, as shown by Mohr et al. (1999). Kigotsi et al. (2017) concluded in a study that stronger storms appear near the eastern border of the country, while the diurnal and seasonal cycles are stronger across the much low-lying areas of the Congo River basin.

However, research attempts investigating the physical processes which control the extensive lightning density over equatorial African region still remain a relatively unexplored topic owing to the lack of in-depth studies here. Also, the mentioned previous studies fail to project a complete knowhow of the mechanisms controlling lightning owing to errors in ascertaining the location or seasonality of lightning as a result of the poor spatiotemporal resolution. Hence, the present attempt uses high-resolution datasets to uncover the distribution of lightning on both seasonal and spatial scales. Next, meteorological features such as moisture availability, instability, orography, aerosols and wind circulation are employed to uncover the underlying mechanism controlling the lightning distribution at seasonal to multidecadal timescales. Attribution of any changes is done through extensive data analysis methods. Lastly, by modelling different atmospheric variables until 2100 under two RCP scenarios (2.6 and 8.5), the study tries to inspect and verify the long-term impacts of a proposed shift in the circulation and moisture transport patterns due to global warming on lightning across Equatorial Congo Basin.

## 2. Datasets

Observations of lightning are taken from the LISs (Lightning Imaging Sensors) placed on the TRMM (Tropical Rainfall Measuring Mission) satellite moving at the rate of 16 orbits daily between 35°N-S (Christian et al. 2003). The sensor is able to detect intra-cloud and cloud to ground lightning flashes with 73±11 % and 93±4 % efficiency during the day and night respectively (Boccippio et al. 2000). Pulses along the 777.4 nm band representing atomic oxygen are monitored using high spatial (5.5 km) and temporal (2 ms) resolution to acquire lightning observations. As soon as the illumination intensity exceeds the background, it gets registered by an illumination pulse, all events within in an integration time period of 2ms is considered to be a single lightning flash. Many past studies have used lightning data using the coarse resolution using a spatio-temporary smoothing of 2.5° and 99 d. Consequently, actual spatio-temporal distribution of lightning properties gives inaccurate estimates. Hence, the present study used actual lightning observations between 1998-2014 from ~95,800 overhead passes. These lightning flash (km$^{-2}$) and lightning radiance values (J m$^{-2}$ steradian$^{-1}$ s$^{-1}$) observations are compiled and averaged monthly at fixed grid resolution of 1° for simplicity. TRMM-LIS sensors have declined in their quality since 2015. thus, the observations are taken for analysis until only 2014

Lightning activity starts due to an extensive charge separation in the mixed phase of convective clouds; however, it requires a sufficient quantity of electrostatic charge to destroy the atmospheric insolation and interact with the ground, which finally lead to widespread socio-economic hazards. Early researchers like Uman (1986), stated that only ~10%-20% of all lightning strikes are strong enough to reach the ground and then cause various hazards. The rest of the lightning are not strong enough and thus remain as intra-cloud in nature. Hence, the overall variation in lightning properties such as its frequency and intensity needs to be investigated in detail as it may help policymakers in avert thus catastrophes under warming scenarios.

Apart from that, the fifth generation of the European Centre for Medium-Range Weather Forecasts (ECMWF) reanalysis, datasets (Hersbach et al. 2020) at $0.25^o$ resolution is utilized in order the unravel the existence of any relationships between the lightning properties and its controlling factors. Corresponding observations of the total aerosol optical thickness (AOT) and also for the black carbon (BC), dust, organic carbon (OC), and sulphate components are accessed in a gridded format from MERRA-2 modelled records from NASA. The reliability of using these products have been explained in many previous research attempts (Buchard et al. 2017). Finally, the multi-decadal projections in lightning frequency and radiance for different urbanization conditions are taken from 5 general circulation models (GCMs) in the Coupled Model Intercomparison Project (CMIP5) archive (https://esgf-node.llnl.gov/search/cmip5/, last access: 13 January 2024) during 1950–2100, using gridded profiles of various meteorological and allied datasets namely: the temperature and humidity profiles along with the aerosol optical depth ( 550 nm). Additional details of all these data inputs have been already depicted in past research (Taylor et al. 2012) and also in the subsequent sections of the study.

## 3. Results and Discussion

**3.1 Spatio-temporal distribution of lightning flashes and radiance**

A plot of the annual mean lightning frequency computed over a span of 17 years from the LIS dataset is depicted (Fig. S1). The global maximum of lightning frequency is observed over the equatorial African region, as expected from past researches. A deep look into the essential characteristics of the equatorial African region reveals the topography to consist of low-lying plains in most of the Congo basin, flanked by highlands in the south and Virunga mountains in the East (Fig. 1a).

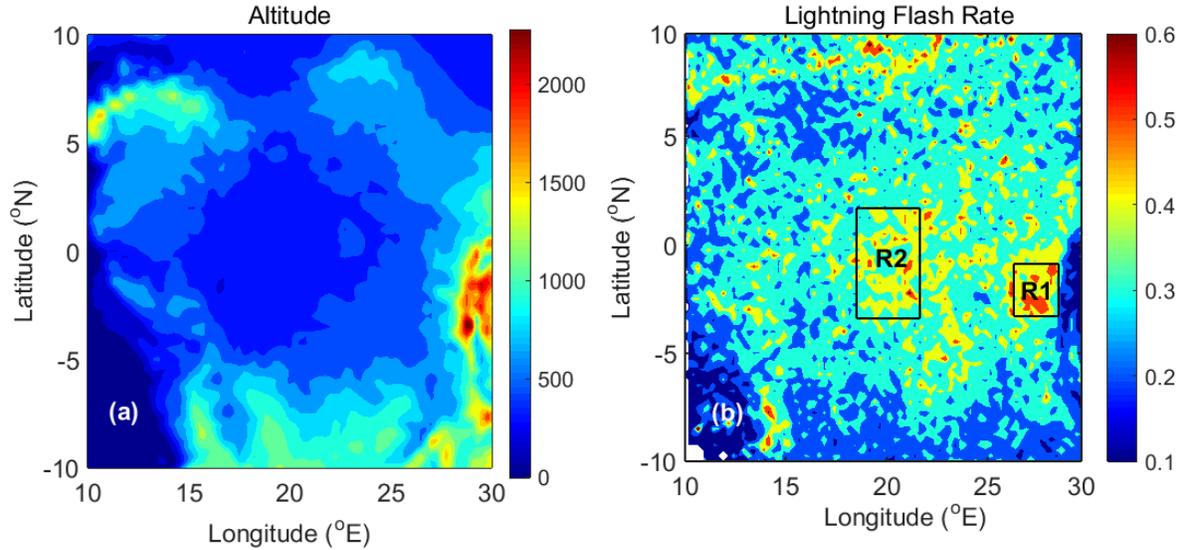

**Fig. 1** (a) Topographical Map of the Congo Basin {10N-10S, 10E-30E}. (b) Annual mean Lightning flash rates per kilometer square at high spatial resolution

The latter is an area important for this study since it is here where the highest values of Lightning Flash Rates (LFR) are observed (Fig. 2b). This peak in LFR values is concentrated along the foothills of the Virunga mountains which is referred to as R1 encompassed by {1.5°S-3°S and 27.5°E-29°E}. Another region R2 {2°N-3°S and 19°E-21°E} is taken such that it encompasses the high values found over the lowest-lying central African region. It may be noted that, R1 and R2 regions are not situated very far from each other (~1000 kms apart), yet significant climatological and geographical differences exist between them. For instance, R2 refers to a hot (~ 26°C), humid (~ 45 kg/m^2), forested and river fed region, near sea level height; while R1 experiences cooler (18°C) and drier (~ 30kg/m^2) conditions along the hilly regions (~1500 meters above sea level). However, on a broader scale, both these regions are part of the equatorial African land region referred to as R3 {10°S-10°S and 10°E-30°E} (Fig. 1b). So, lightning properties such as its frequency and its intensity have been studied across the regions R1, R2 and R3, averged per square kilometer.

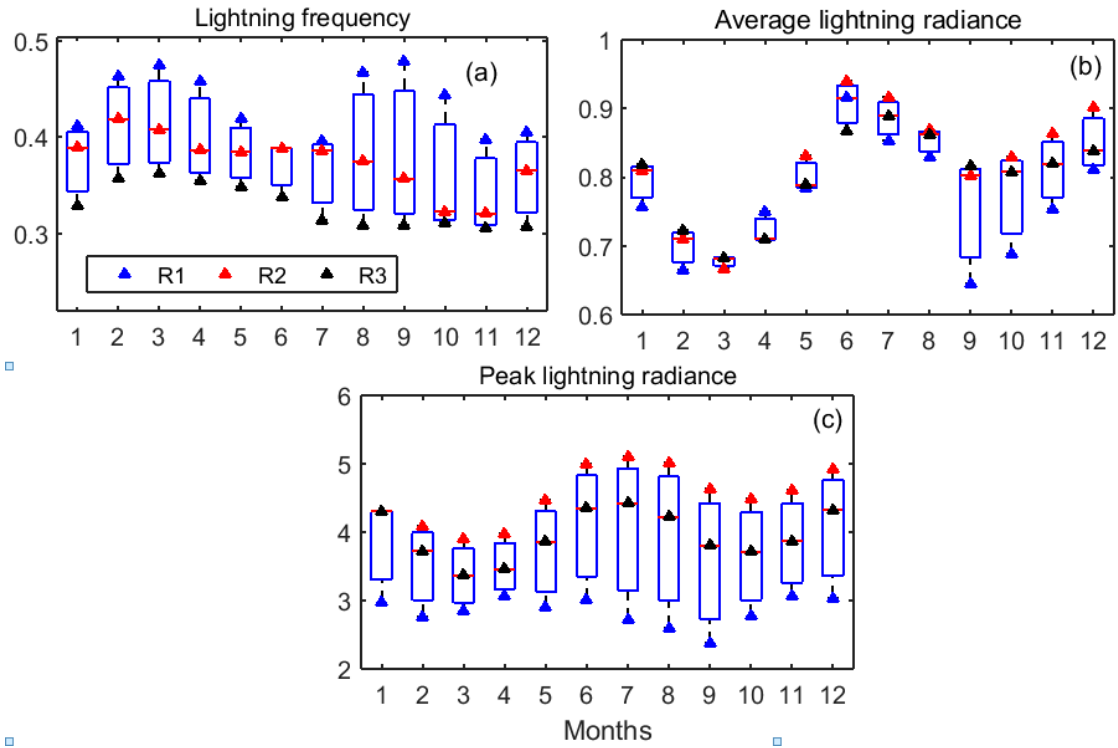

**Fig. 2** Seasonal Variations in Different Lightning Properties – (a) LFR, (b) Average lightning radiance, (c) peak lightning radiance

Looking into the seasonal trends of lightning features within the study regions, it is first observed that LFR values peak during the equinoctial months (February – March and August-September) (Fig. 2a). Also, LFR values are significantly higher in R1 than in R2 across the year. This can be due to the orographic assistance that lifts air parcels along the foothills of mountain ranges in R1. In addition, LFR shows a drastic seasonal decline along the boreal summer months (June-July) in both regions, akin to previous studies (Soula and Chauzy 2001). However, Lightning radiance follows an opposite seasonal trend where the average radiance peaks in June-July and shows strikingly low values over February-March and September (Fig. 2b). Also, region R2 shows unprecedented high values of lightning radiance in June. The peak radiance - defined by the 90th percentile of lightning radiances – also shows a similar disparity between the seasons and regions, with R2 consistently showing higher values of peak radiance than R1, especially in June (Fig. 2c). In addition, higher LFR values are seen in R1 and R2 than in R3, but in the case of lightning radiance, the highest values are observed over R2, followed by R1 and R3.

To understand the reason for the observed seasonal disparity among the lightning properties, it is essential to understand how lightning occurs. Lightning is caused by the electrical breakdown of air, due to the separation of charges within a cloud. A higher concentration of cloud particles in both ice and liquid cause a higher magnitude of charges to be separated, resulting in lightning flashes with higher radiance values (Takahashi 1978; Mansell and Zigler 2013). Hence, the present study attempts to explain the seasonal and annual variations in lightning features based on the variations in moisture content and cloud properties and the various factors that contribute to it. Accordingly, a detailed investigation is performed on the lightning properties (Fig. 3a-c) along with a set of various probable controlling factors as suggested by previous attempts, such as instability (CAPE), wind speeds, surface conditions, moisture availability and the cloud and aerosol parameters.

**3.2 Spatio-temporal distribution of atmospheric variables impacting lightning**

Over R1 and R2, CAPE follows the seasonal path of the incoming Solar radiation over R3, with the highest values during the equinoctial months and lower ones during the rest of the year (Fig. 3d). R2 shows higher values of CAPE than R1 primarily due to the presence of more moisture content (Fig. 3j) and higher surface heating which in turn arises from its lower elevation (∼ 300 m) in comparison to R1 which is situated at a greater altitude of 1500 meters. This is also supported by the surface temperature and pressure values over R2 in comparison to R1 (Fig. 3(e-f)). Next, as cloud formation requires the presence of strong convective instability, as given by CAPE, hence this parameter follows the seasonal distribution of LFR very well. However, it may be noted that orography also plays a significant role in generating lightning flashes in R1 due to its location on the windward face of the Virunga mountains. Next, in agreement with the movement of ITCZ, total cloud cover peaks over the equator in the equinoctial months and reduces during the austral summer months (Fig. 3g). But interestingly, the cloud liquid water content increases sharply, during June-July especially in R2 (Fig. 3i). Thus, it follows that some external influence from large-scale dynamics can also have a strong role towards such phenomena during the months of June and July, as this is also supported by the high values of mid-cloud cover during the same time of the year (Fig. S2d).

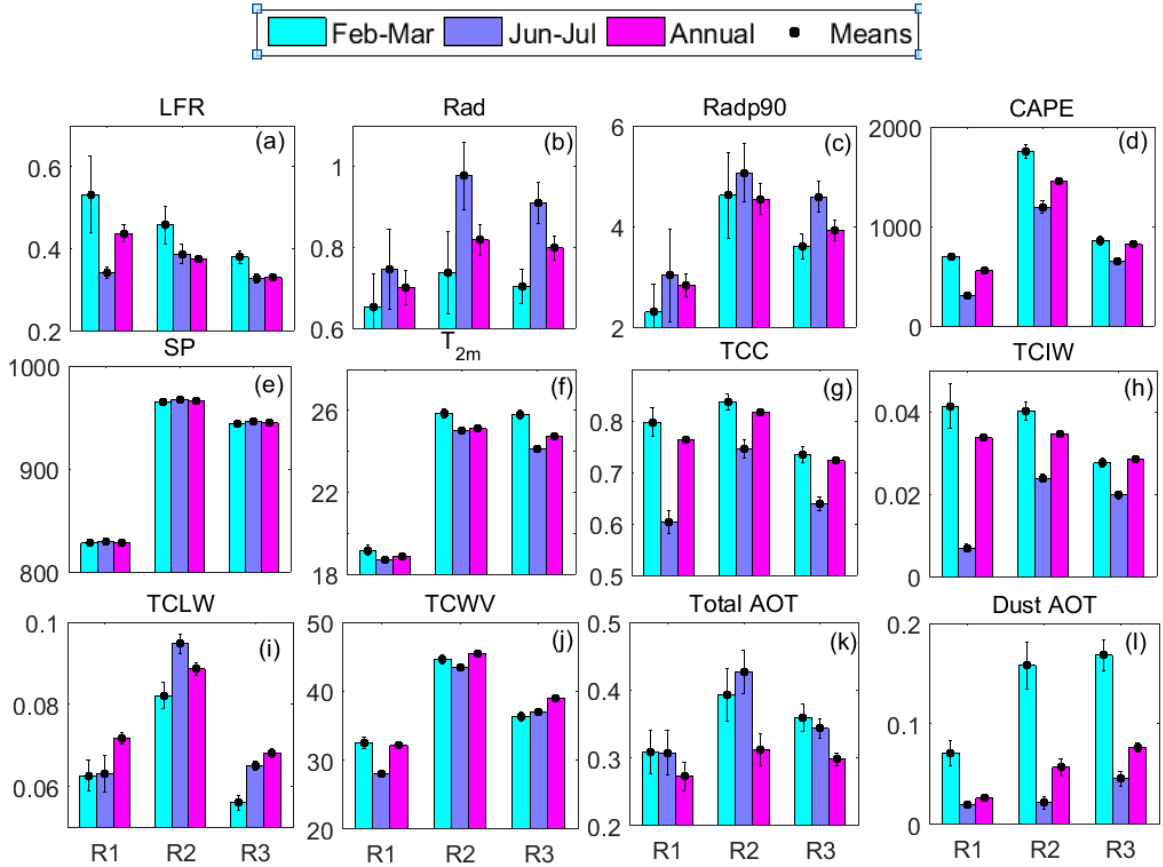

**Fig. 3** Climatological average values of lightning properties and its various controlling factors during February and March, June and July and throughout the year over Regions R1, R2 and R3 namely: (a) Lightning flash rate, (b) Average lightning radiance, (c) Peak lightning radiance, (d) CAPE, (e) Surface pressure, (f) 2 meter air temperature, (g) Total cloud cover, (h) Total cloud ice water, (i) Total cloud liquid water, (j) Total column water vapour, (k) Total aerosol optical thickness, (l) Dust aerosol optical thickness, Standard deviations are represented as error bars

Next, looking into the wind patterns, it is found that the winds in R1 and R2 are predominantly towards the North and East directions (Fig. 5). However, owing to the position of ITCZ, near-surface wind speeds over R1 and R2 are stronger during June than in February. However, the prevailing winds are complemented by orographic convection over R1, thus showing high TCC values despite lower CAPE values. R1 and R3 show higher values of TCC in February than in June-July, as also supported by the cloud ice water content values (Fig. 3h). However, region R2 experiences frequent Congo basin cells (CBC) (Longandjo and Rouault 2020) and low-level westerlies (LLW) (Nicholson and Grist 2003) during June, and this results in stronger convection and lightning over R2 in contrast to R1 and R3.

Next, the analysis progresses to observe the impact of aerosols. AOD assumes higher values in the month of June over R2 than during the rest of the year (Fig. 3k). Dust and OCs dominate the aerosol composition with lower contributions from sulphates and BC (Fig. S2l). However, it is to be noted that the concentration of dust aerosols is significantly higher in February than that in June (Fig 3l). This can possibly be due to them being carried by the winds converging into the ITCZ positioned over the central African region during the equinoctial months (Collier and Hughes, 2011). Dust aerosols are known to give rise to ice crystals and, hence, align with the seasonal trends of TCIW (Total Column Ice Water) and LFR in region R1. On the contrary, a lower concentration of dust aerosols is seen in the boreal summer owing to the absence of any convergence into the study region, as the ITCZ has migrated towards North Africa. The sulphate aerosols show high concentrations over R1 throughout the year, which can result from high tectonic activity in this region. Sulphates are known to be good condensation nuclei and thus have an equal role as dust aerosols in the formation of higher quantities of cloud particles in R1. However, sulphates, OC (Organic Carbon) and BC (Black Carbon) are seen in more numbers in June than in February. This can be attributed to increased rainfall in February, as it can wash away aerosols in the air by both rainout and washout processes. Thus, the higher concentrations of aerosols may also partially contribute towards the stronger radiance observations at R2 during the boreal summer months.

### 3.3 Long-term trends of lightning properties and its controlling factors

The trends in lightning properties and their causative factors across 17 years (1998 – 2014) are now examined. Over R1, a rise in the lightning flash rate is seen (0.5%/year) (Fig. S3). CAPE and AOD also show a slight increase despite no other parameters showing similar changes (Fig. 4a). To understand the reason for the observed disparity among the trends, it must be noted that higher CAPE over a certain region causes more turbulence and stronger updraft speeds, which in the presence of more nucleating agents leads to more ice-graupel collisions, and a higher number of lightning flashes. However, such trends are not seen in R2, and thus, this region does not experience a corresponding rise in LFR.

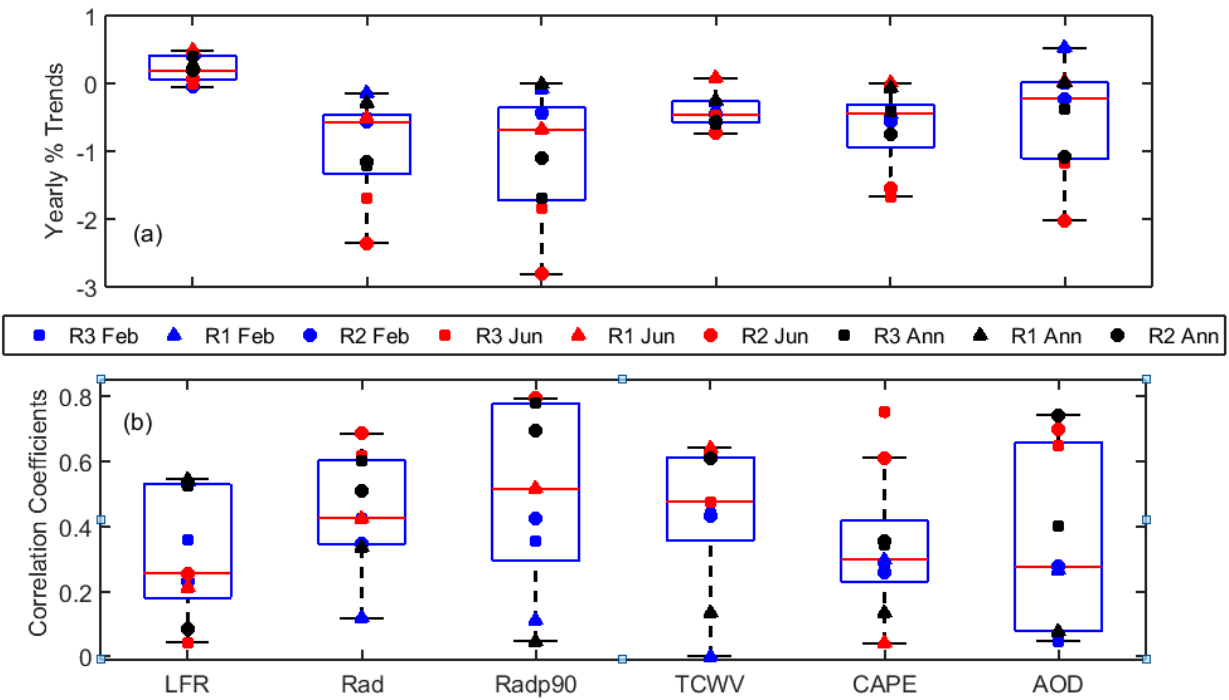

**Fig. 4** (a) Percentage trends in the Lightning properties and its controlling factors, namely Lightning flash rate, Average lightning radiance, Peak lightning radiance, Total column water vapor, CAPE, and Aerosol Optical depth (b) Autocorrelation (corrrelation with time) coefficients of the observed trends

On the other hand, the radiance of lightning flashes is determined by the amount of charge transferred during collisions between the mixed phase hydrometeors and this, in turn, is controlled by the number and size of those particles. This makes it vital to understand the relative impact of aerosols and cloud water content in controlling the lightning intensity. Now, a large decrease of -2%/year in lightning radiance is observed over R2 with high values of its temporal autocorrelation during June and July (Fig. 4a-b and Fig. S4). As the radiance of lightning is dependent on CAPE and AOD, the observed weakening in lightning intensity matches with a corresponding decrease in these parameters as well. Among the aerosol species, dust aerosols, organic carbon and sulphates show significant decreasing trends (-1%/year). Apart from this, near-surface winds (u100 and v100) and cloud contents (TCLW and TCIW) also show a strong decline over R2 with a reasonable autocorrelation with time, and this hints towards a possible decline in moisture transport into this region during the last few years. In contrast, R1 depicts an increase in the values of only CAPE and AOD but not the prevailing moisture or cloud content. Consequently, the number of ice droplets and graupel did not change, leading to an insignificant change in the radiance values.

### 3.4 Wind patterns over tropical west Africa

The wind circulation patterns are studied for the months of February-March and June-July to examine whether large-scale dynamics and moisture transport play any role behind the observed weakening in lightning radiance over R2. It is seen that the principal direction of moisture-laden winds from the Atlantic Ocean shifts northwards from equatorial Africa in February-March to the North African Sahel region during June-July due to the seasonal migration of ITCZ (Fig. 5).

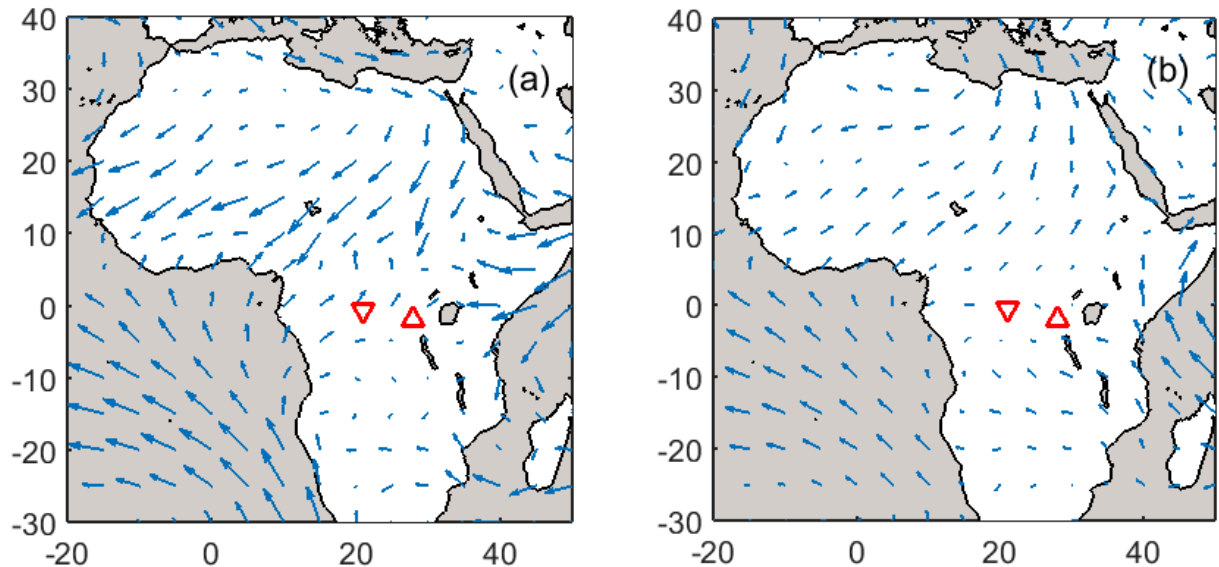

**Fig. 5** Composite Analysis of the wind circulation patterns at 950 hPa during (a) February-March (b) June-July. R1 and R2 are marked by the upward and downward triangle, respectively

      The winds blowing over the western equatorial region of Africa have already been discussed explicitly in past research (Pokam et al. 2014). Winds blowing into the Congo Basin from the Atlantic Ocean and are stronger during the local summer season. Meanwhile, in June-July, the trade winds are now directed northwards to a convergence zone in the Sahel region (Fig. 5). However, some of these cross-equatorial winds into the Congo Basin persist and harbour moisture into the R2 region, thereby playing an important role behind the unseasonal lightning occurrences over this region. The same cannot be said for R1 as this region experiences much weaker winds and lower moisture influx compared to R2 (Fig 3(j), Fig S2(f,g)). As mentioned earlier in the study, the zonal and meridional components of wind have weakened drastically over R2 in June across the years. Changes in the pattern of the winds over R2, can thus directly affect the flow of moisture and aerosols from the Atlantic Ocean.

      Hence, in order to in order to understand the sensitivity of the moisture influx over the study region, the seasonal changes in the moisture flux are examined with respect to the local moisture content during boreal spring (February-March) and boreal summer (June-July) months. Here, moisture flux has been defined as the difference between the amount of moisture flowing into and out of the area. To get rid of seasonality issues, the moisture flux values have been normalised for both of the regions. Now in typical cases, a gradual but prominent strengthening in Moisture Flux is expected to bring about a parallel rise in TCWV as well. However, the extent of this dependence is found to be less over R1, which can be attributed to the weakening and drying of winds before reaching this region (Fig. S5). However, the situation improves over R2 as the Moisture Flux follow the TCWV variation very clearly, with the inter-cluster differences being much larger than their mutual standard deviations, along with higher correlation coefficient values. So, the large-scale processes involving moisture transport are found to be a major player behind the lightning occurrences over the study region and, hence, need to be further investigated.

      Next, it is attempted to investigate that which factor among moisture availability, thermodynamic instability and aerosol nucleation plays the most important role behind lightning frequency and radiance (Fig. S6). For this, the magnitudes of all these components are normalised with respect to their mean and standard deviations to obtain uniformity. Next, the lightning data of 17 years is sorted into two non-overlapping clusters of 8 years each for low and high lightning for three different cases, namely FM, JJ and Annual and also for the two lightning properties (LFR and Rad). The mean and standard deviation of each of the two clusters are then plotted with respect to each input parameter. Parameters showing the most prominent contrast between the two clusters are considered to play the most decisive role in each spatiotemporal scenario. Over R1, AOD shows the most prominent separation between the clusters irrespective of seasons, while CAPE shows the weakest inter-cluster deviations due to the absence of surface and boundary layer convection. The impact of MF is also negligible, but it takes second place in case of radiance variations, in the R1 region. Over R2, the importance of MF becomes primary in almost all cases, while the effects become stronger over Radiance during JJ. This is followed by AOD, which plays a secondary impact here.

So, it implies that in the case of R2 being closer to the oceans, the impact of the advected MF remains dominant irrespective of the aerosol or instability influences, as also hypothesised earlier.

Next, a multi-linear regression analysis is carried out to quantify the relative contribution of each input parameter (moisture flux, CAPE and AOD) towards the variation of lighting properties over R1 and R2 for both the seasons (Fig. S7). Over R1, both lightning properties are influenced more strongly by AOD than the Moisture flux. This can be attributed to the presence of higher amounts of sulphates and dust over R1 which in turn contributes largely to lightning according to previous research regimes. Over R2, moisture flux and aerosol content are seen to be playing an almost equal role behind lighting radiance. Aerosols, however, play a more important role than Moisture flux or CAPE in case of lightning flash rates. Nevertheless, an opposite behaviour is followed in peak radiance values, where moisture flux has a much stronger contribution. This suggests that a long-term trend in moisture flux will have the strongest impact on peak radiance followed by the average radiance and then by the frequency of lightning.

### 3.5 Change of wind transport regime

From previous sections, it has become evident that incoming moisture from the Atlantic greatly influences peak lightning over the study region. Hence, it was hypothesised that the observed multidecadal decline in lightning properties over the study region could also be related to a concurrent variation in the moisture influx. This can also be well supported by the wind circulation plots in previous sections (Fig 5) depicting the prevalence of south-westerlies from the Atlantic rushing into the R3 region.

Typically, a contrast between the temperatures over land and the adjacent Gulf of Gabon creates a breeze that flows from the Gulf towards the heated land, carrying moisture. However, it is very interesting to note that both the Sahel region in Northwest Africa and the Congo Basin share a common source of moisture-laden winds in June and July when the ITCZ is stationed above the Sahel region (Fig. 6j). Hence, it can now be proposed that any anomalous heating over either the Sahel region or the Congo basin would draw in the winds towards the same region due to a corresponding decrease in the surface pressure. Increased winds into the heated region would lead to an increase in moisture flux, inducing a deficit in the winds and the moisture flux in the other region, leading to a probable weakening in lightning intensity.

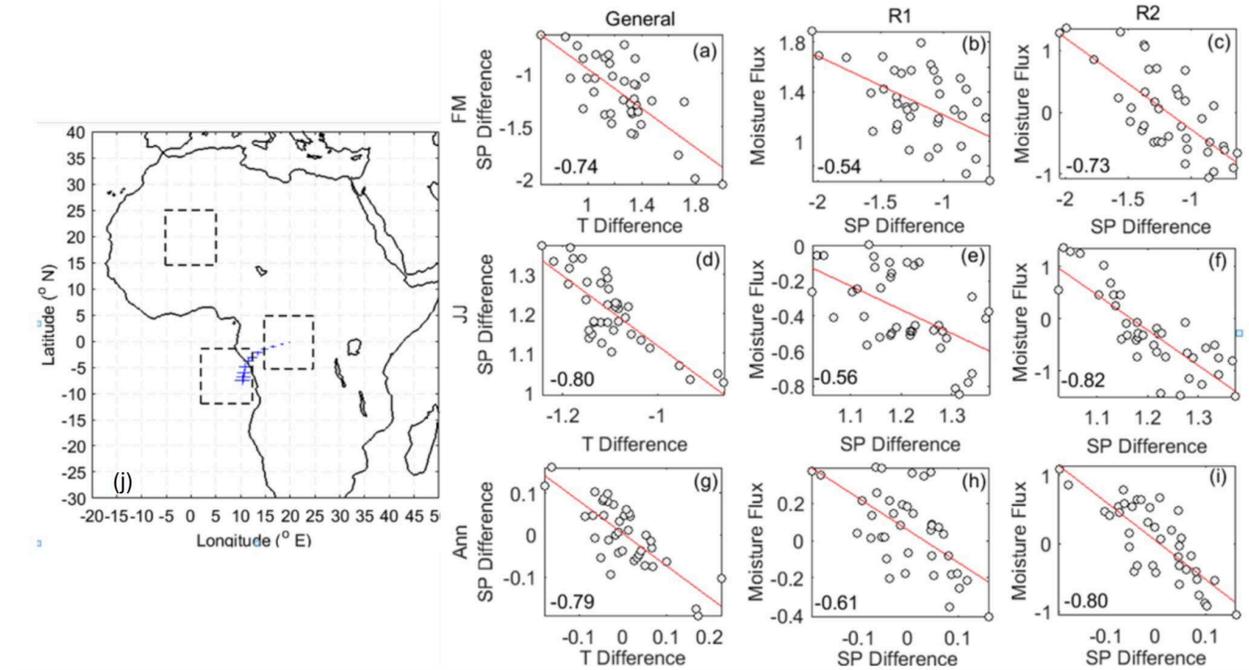

**Fig. 6** Correlation between various spatial gradients across 1979-2017 between (a, d,g) dCS of surface pressure and dCS of temperature for the entire study region R3, (b, e, h) dCS of surface pressure and moisture flux into R1 and (c, f, f) same for R2 during three time periods eg February-March (a-c), June-July (d-f) and Annual (g-i) shown in [Rightf and the Schematic map showing the utilized land and sea region [Left] (j). Error as standard devaitation of the predicted backtracking path of incoming water vapour into the R1 and R2 are marked in blue

To understand the validity of such a hypothesis, three box regions are first considered. The first one is centred around R2 as it experiences the strongest anomalous weakening in lightning. Next, the second box is taken over the Sahel region in such a way that it would be of a similar area to the first and then would be of comparable distance from the sea as R2 and finally would correspond to the centre of wind convergence during JJ as observed from the circulation patterns. This region has been taken as a box encompassing {15-25deg N and 5E-5W} in latitude and longitude bounds. The Sahel region has seen large scale changes in its weather patterns in recent years, especially with the widening / northward shift of Hadley Cell. Since this region, being close to the shore, can act as a sink of moisture, is analysed. Finally, the third box is selected over the Atlantic Ocean centred around that particular grid point which corresponds to the most probable source of moisture from the sea regions as also validated from the long-term histograms of the 3-5 day back trajectories at 950 hPa level using the Hysplit back trajectory analysis during the last week of June and the first week of July between 2006-2014 (https://www.ready.noaa.gov/hypub-bin/trajtype.pl?runtype=archive).

Next, to test whether the proposed hypothesis holds true, the study tracks the "difference in heating" between the Congo Basin and the Atlantic Ocean, as well as that between Sahel and the Atlantic Ocean, over a set of 38 years (1979-2017) from ERA-5 reanalysis data. Since the Atlantic Ocean is a common term in the above calculations, the net difference is essentially the difference in heating between the two land regions. This difference for any parameter between the Congo Basin and the Sahel region will be henceforth referred to as dCS of that parameter.

The Sahel region has been observed to warm at a faster rate than the study region during 1998-2014. Hence, it is expected to weaken the land-sea contrast between the Atlantic and the study region, thereby leading to much weaker moisture influx and cloud water content and, hence, much weaker lightning over R2. To check this, the difference between Sahel and each of the regions (R1, R2 and R3) is calculated for each parameter and plotted against each other for the total available data span of 38 years (Fig. 6a-i). During June and July, a strong correlation is observed between the dCS of surface pressure and temperature (-0.80). This explains how an decrease in the dCS of temperature, would correspond to a stronger warming over the Sahel region, and hence should cause an increase in the dCS of surface pressure. Next, looking into the correspondence between the dCS of surface pressure and dCS of moisture flux, a negative correlation is seen over both the regions, with higher values over R2 (-0.82), followed by R1 (-0.56). This substantiates the possible mechanism linking the observed decrease in dCS of surface pressure with the dampened land-sea contrast, which in turn has weakened the moisture influx over R3 as proposed. Notably, similar trends are also seen during February-March and annually, further supporting the proposed mechanism.

To further validate the proposed mechanisms between Sahel and R3, anomalies of the lightning properties, moisture fluxes and dCS of temperature and surface pressure are compared between two different years from successive decades between 1998-2014 (Fig. S8). During the year 2000, higher temperatures are observed over R3, which leads to higher moisture influx there and, consequently, higher radiance values. Contrastingly, in the year 2010, higher temperatures over the Sahel than R3 led to a decrease in moisture flow into the R3 region, causing lower radiance values. Notably, the peak lightning radiance shows an even higher difference in the values of the two years. Also, since R2 is closer to the Atlantic Ocean than R1, the observed weakening in lightning is more prominent over R2 than R1. However, since LFR is not primarily dependent on MF, the disparities between the two years are not visible for this parameter.

**3.6 Future projections:**

It has been demonstrated in the previous sections how the observed warming over the Sahel region has led to the redistribution of moisture and weakened lightning over the global lightning hotspot in recent decades. However, it is now necessary to confirm whether the same mechanism will continue to be in effect in the coming years, assuming various global warming scenarios. The utility of such a study becomes even more imminent considering the projected warming of the subtropics in response to global warming, as also hinted at in many current research attempts. In the recent past, many studies have attempted to utilise global climate model simulations to project lightning under anthropogenic climate change until the end of the current century (Chen et al. 2021; Michibata et al. 2024; Asfur et al. 2020). Hence, lightning trends and their various controlling factors are analysed and depicted across a 150-year time span (1950 - 2100) using global climate model simulations similar to previous attempts (Chakraborty et al. 2019). The trends in lightning properties have been ascertained by passing the controlling factors through a multilinear regression model with coefficients which were already obtained in the previous sections. These projections are drawn under two diverging greenhouse emission scenarios, namely the RCP 2.6 and RCP 8.5 pathways, representing the increase in the net radiative forcing at the top of the atmosphere due to the corresponding increase in CO2 levels. Observations from these warming scenarios will help to extrapolate the bearing of the changing large-scale dynamics on lightning over the Equatorial African region.

The next step is to search for global climate models which provide reliable estimates of temperature, humidity and wind profiles daily along with the monthly average aerosol concentrations. The daily meteorological profiles are utilised to run standard parcel theory simulations to derive the monthly averaged values of CAPE and MF for both scenarios as already done in previous attempts (Chakraborty 2021). A set of 5 GCMs from the CMIP5 inventory at r1i1p1 realisation are found suitable enough as they provided all the required meteorological parameters, and a brief description of those is available in Table 1 (Supplementary). Next, it is required to select well-performing model(s) which provide the most reliable estimates of all the parameters for the multidecadal lightning projections. Thus, the atmospheric parameters are utilized for the period 1950 - 2005, and the model providing the least deviation with respect to the observed datasets from ERA5 reanalysis is selected. As evident from Fig. S9, only the NOR-ESM model is found to be the best-performing model for further simulations, owing to its acceptable correlation and lesser deviations from the ERA5 data. Thus, this model has been selected for further analysis until 2100.

Consequently, the 150-year variations of the dCS of Temperature and Pressure are observed in normalized form for both greenhouse warming scenarios in the form of 10-year averaged intervals (Fig. 7). In the case of temperature, the RCP 8.5 scenario depicts a drastic decline, implying that the Sahel region would continue to heat up faster across the century. However, a dampened decrease is found in the RCP 2.6 scenario, as expected from previous research results. Surface pressure is directly dependent upon the surface temperature. Hence, in the 8.5 scenario, the dCS increases prominently indicating that the Sahel will experience lower surface pressures than the R2 region, which further continues to decrease until the end of the century. But, in the 2.6 scenario, this increase is much weaker than the extreme warming scenario.

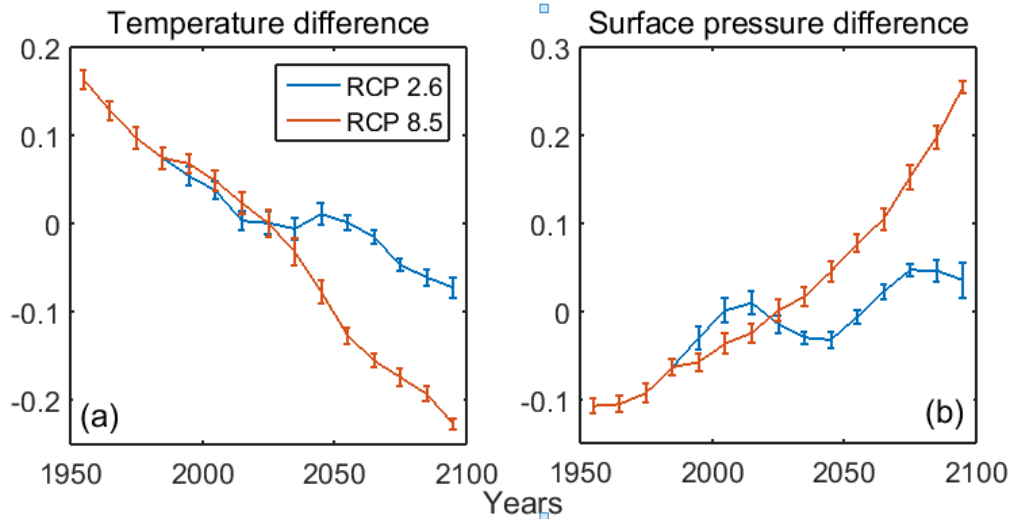

**Fig. 7** Multi-decadal trends of dCS for (a) Surface temperature and (b) Surface pressure from 1950 to 2100 between Congo Basin and Sahel region for RCP 2.6 and 8.5 scenarios

Next, it is attempted to depict the impact of the aforementioned trends on lightning and its controlling factors during the June-July period (Fig. 8). First, as expected, the moisture flux depicts a decline over both R1 and R2, though the degree of weakening in the moisture influx is much slower in RCP 2.6 than in the RCP 8.5 scenario. But very interestingly, the RCP 2.6 cases depict an increase in the values after 2050. This can be explained assuming that if there is no differentially stronger heating over Sahel, then it would not have any impact on the oceanic moisture distribution, however, the gradual warming would also lead to more moisture holding capacity in the air resulting in a higher moisture flux in the course of time. The timing of the observed diversion from the expected trends can be attributed to climate inertia and natural latency in multidecadal scales. Next, in the case of aerosols, under the 8.5 emissions scenario, AOD rises initially, but then, the values stagnate and remain constant in both regions. This could be due to the fact that as the differential surface pressure increases, the Congo Basin results in weaker winds, leading to reduced aerosol loading across R3. In the case of CAPE, the RCP 8.5 scenario, depicts a linear increase, which is a direct result of stronger surface forcing due to increased urbanisation and global warming, while the trends are much weaker in the RCP 2.6 case as expected.

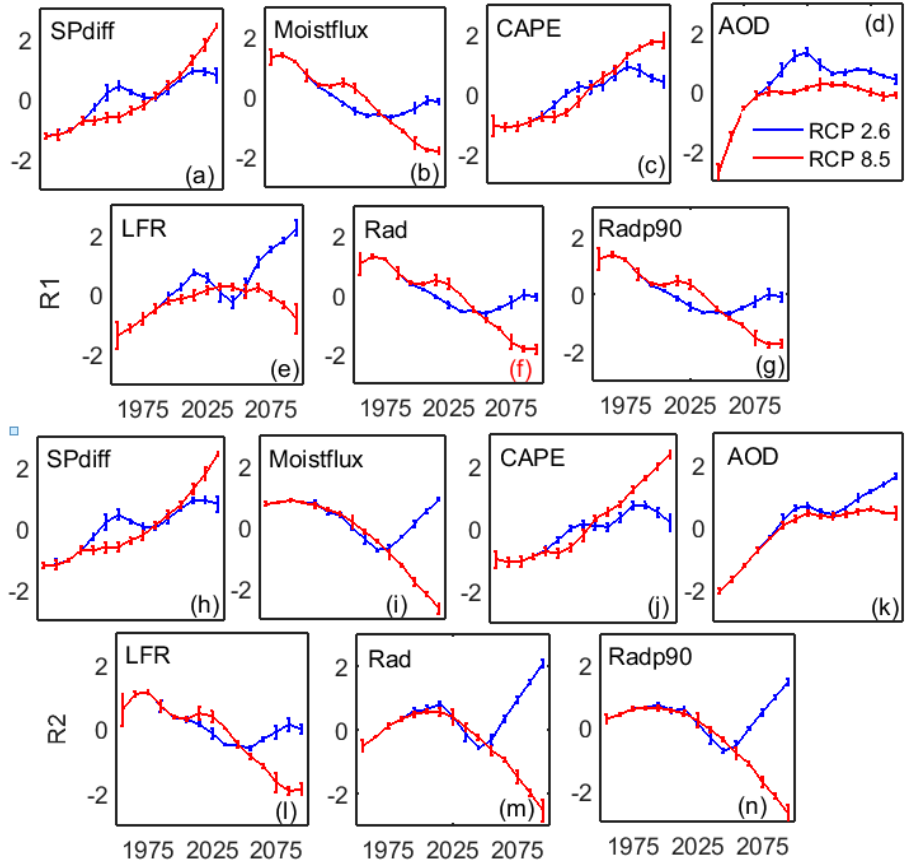

**Fig. 8** Climatic projections of lightning properties and their controlling factors namely: Surface pressure gradient (a, h), Moisture flux (b, i), CAPE (c, j), AOD (d, k), Lightning frequency (e, l), Average lighting radiance (f, m) and Peak lightning radiance (g, n) over (a-g) R1 and (h-n) R2 from 1950 to 2100 for June-July months

Lightning flash rates over R1 exhibit an overall decrease in the RCP 8.5 scenario. This can be attributed to the combined effects of moisture, CAPE and aerosols. While CAPE and aerosols increase in the first half of the study period, aerosol values stagnate throughout the latter part. More importantly, moisture flux values are ever decreasing, leading to a decrease in the flash rates. In R2, however, the decrease in the LFR is much more prominent, possibly due to the correspondingly stronger decrease in moisture flux values. Nevertheless, in the RCP 2.6 scenario, over R1 and R2, both moisture flux and aerosol loading increases after 2050. This is due to the absence of the Congo-Sahel differential heating episode in a gradually warming climate which in turn would lead to the intensification of incoming southwesterlies into R3; hence, LFR shows a prominent increase in R1 in the lower forcing scenario. Next, coming to the intensity of lightning, the average and peak radiance values follow the trends for moisture flux very closely in both regions due to the strongest impact of the latter on the former. Hence, they depict contrasting responses in both scenarios. However, in line with the present trends of global warming, it can be safely projected that the lightning intensities are going to weaken drastically over R2 and also slightly over R1 by 2100.

In contrast to the June-July period, the ITCZ is positioned over the Congo Basin in the February-March months, and this is expected to act as an ample source of moisture for lightning over R3. Hence, the hypothesised theory of differential heating should cast little or no impact in this case (Fig. S10). Consequently, R1 depicts a weak to moderate increase in the lightning features due to the expected moistening of these areas in a warming scenario. However, over R2, the situation is slightly different. A visible decline is seen in the moisture flux and lightning properties, like in June-July months in the case of the RCP 8.5 scenarios, indicating that regions closer to the coasts still depict an impact due to the prevalent differential land heating between the Sahel and Congo region. However, the degree of such changes is much lower due to the dominant presence of the ITCZ. In the case of the lower urbanisation scenario, the lightning properties remain unchanged with time with some minor multidecadal variations.

Finally, it is essential to highlight the significance of the projected changes across all the trends. Accordingly, the percentage change in the values of lightning properties between two decades 2091-2100 and 2011-2020 is calculated (Fig. 9). Starting from LFR under the RCP 8.5 scenario, the R1 subregion experiences a decline by more than 20 % in June-July. In contrast, the same region experiences an increase of ~20% in February-March, hence the annual LFR change remains weak. However, under the same scenario, R2 experiences a prominent annual decline (~-25%) which is again stronger during June-July (~-30%) than in February-March (~-15%). Next, under the 2.6 forcing scenario, there is an increase in LFR over the R1 subregion, which is stronger during June-July and weaker in February-March, leading to a slight LFR increase annually. But over R2, LFR hardly shows any prominent variation.

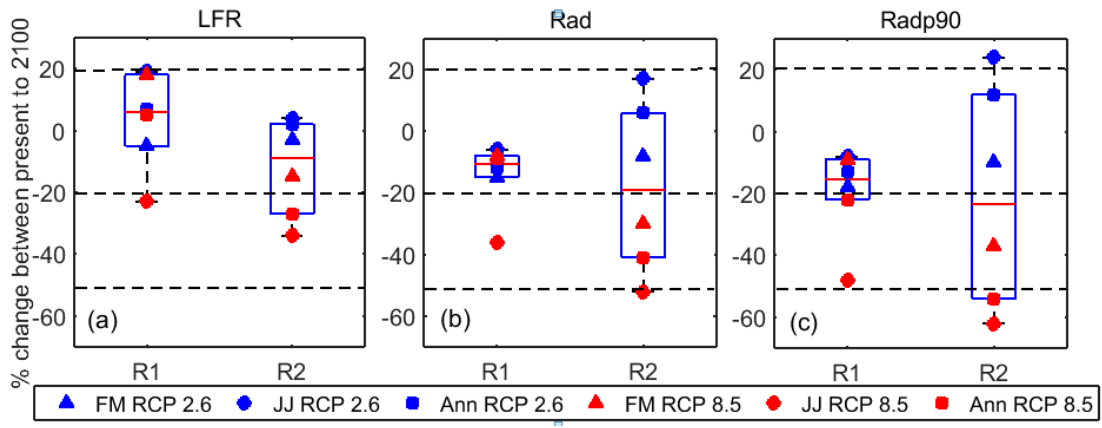

**Fig. 9** Percentage trends of (a) Lightning flash rate, (b) Average lightning radiance, (c) Peak lightning radiance values over R1 and R2 with respect to the 2011-2020 decade using RCP 2.6 and 8.5 scenarios during the February-March, June-July and Annual periods. Horizontal lines across the plots indicate the respective % changes from the 2011-2020 values

On the other hand, lightning radiance experiences strong changes over R2, in all cases, as expected from the multidecadal trends of its controlling factors. Under the RCP 8.5 scenario, the lightning radiance values decrease significantly, irrespective of the season. By 2100, the radiance values are expected to weaken by ~50% during June and July. This is followed by a ~40% decline in annual values and ~30% in February-March, even in the presence of ITCZ over the region. Now coming to R1 in this scenario, the radiance continues to exhibit a strong decline (~-30%) in June-July, while the February-March months and the annual cases depict only a slight decrease (~-10%) as expected. In the lesser forcing scenario, R1 shows a weak decrease across the seasons and annually (~ -10% to -20%). In R2, however, the radiance values increase by nearly 20% during the June-July months, while these changes are much subdued in the annual scale. Finally, the peak radiance closely follows the radiance trends, but only the observed variability is much higher. In the case of the RCP 8.5 scenario, peak radiance values fall sharply during June-July months (~-60% in R2 and ~-45% in R1), and this is also reflected from the annual trends (~-20% in R1 and ~-50% in R2). However, in the RCP 2.6 scenario, the peak radiance values increase significantly in R2 during the June-July months and also at an annual scale (~20%), while relatively, much weaker trends are observed over R1.

## 4. Conclusions

The global tropics are home to a majority of lightning flashes which pose great damage to infrastructure and life owing to the uncertainty in their position and timing. Various natural and anthropogenic factors modulate lightning, though the scalability of such impacts varies spatiotemporally across past research. Hence studies on understanding the physical mechanisms leading to lightning have now become an emerging topic of research in a gradually warming climate where the weather extremes are all projected to intensify. The Congo River basin and the adjoining slopes of the Virunga mountains receive the most prominent share of lightning along the global tropics due to a combination of several factors such as the bi-yearly passage of ITCZ, moisture convergences from Atlantic and aerosol nucleation impacts. However, a detailed study to delineate the importance of each of these factors is necessary owing to the poor resolution of the data and scarcity of studies in this field. Hence, the present study uses high-resolution datasets to uncover the spatio-temporal distribution of lightning occurrences on sub-seasonal scales and also tries to uncover the underlying mechanisms controlling the lightning distribution using detailed statistical analysis. Besides, an attempt is also made to regress the obtained understanding about the lightning properties into climate models for generating multidecadal forecasts of lightning frequency and intensity.

A high-resolution spatial distribution of lightning over the Congo Basin revealed two prominent patches: one along the slopes of the Virunga Mountain Range (R1) and the other inside the Congo rain forests (R2). The frequency of lightning is observed to be higher over R1 than R2 while the opposite is seen in case of the lightning radiance. Regarding seasonal distribution, high LFR values are observed over R1 during the equinoctial months, which is found to be associated with the impact of orographic assistance and ITCZ influence, followed by aerosols. On the other hand, the average and peak radiance values are much stronger in the boreal summer months (June-July), with unexpectedly high cloud liquid and ice water content, especially over R2. This can be attributed to a moisture convergence component from the adjoining Atlantic Ocean during the local dry season, as also evidenced by the wind circulation patterns.

Next, the 17-year trends of lightning properties over the Congo basin reveal a steady increase in LFR (~0.5%/year) over R1 during February-March while a significant weakening in average and peak radiance (~-1.5%/year) is depicted over R2 during June-July. The observed increase in LFR over R1 can be attributed to an increase in dust loading over R1 region, causing a greater number of cloud particles and, thus, higher LFR. However, the above-mentioned weakening in lightning intensity is unprecedented in a warming climate and it is found to be caused by a decrease in the incoming moisture flux from the Atlantic towards the Congo Basin. A detailed statistical investigation (involving temporal clustering, multi-linear regression and case studies) inferred that the observed reduction in the moisture content is driven by the weakening of south-westerlies during the June-July months which in turn is caused by the accelerated heating of Sahel region under climate change, leading to a stronger land-sea contrast towards the north of the study region and this eventually weakens the share of moisture transport through Gulf of Gabon into the Congo Basin region.

Finally, it is attempted to provide multidecadal projections of the lightning properties under various urbanization scenarios using Global model simulations. Statistical intercomparison between the historical data and the reanalysis revealed the suitability of NorESM-1ME model for performing the projection analysis. The RCP 8.5 scenario depicted a drastic decline in lightning radiance over R2 during June-July (~-50%) due to the projected rise in moisture deficit from the Atlantic towards the Congo basin. However, a sudden increase in lightning radiance is observed under the 2.6 scenario (20%), which may be likely due to the increased moisture flux into Congo basin due to the absence of accelerated heating over Sahel in a much slowly warming climate. The peak radiance shows even stronger impacts of the proposed moisture deficit hypothesis as it depicts a much larger weakening during June-July (~-60%) under the RCP 8.5 and a prominent increase in the RCP 2.6 case. Notwithstanding, the LFR values show a large increase in the RCP 2.6 scenario during February-March (~20%), whereas, a strong decline is observed in the RCP 8.5 scenario during the June and July months over R2 (~-30%), and this can again be attributed to the influence of the proposed moisture flux dynamics.

The present research emerges as the first ever study which uses high resolution observations to resolve the fine spatial heterogeneities in lightning density over the Congo River Basin and then it also attempts to identify the different physical mechanisms controlling the lightning genesis over R1 and R2 separately. Further, this study also presents two counter-intuitive results such as the seasonal intensification of lightning along the dry months of June-July and also the observed weakening in lightning intensity in a warming climate and it also attempts to explain such behaviours. The obtained understanding from this study is supported by adequate statistical and physical basis and hence can be applied to similar tropical locations, which also depict an anomalous reduction in extreme weather trends in a warming climate. However, this attempt still has a couple of shortcomings. First, the utilized data span of lightning is only 17 years (1998-2014). Yet, satellite-based lightning observations from the International Space Station borne LIS have not been utilized owing to their even smaller data spans (2017-2023) and their different declination angles and revisit times compared to the TRMM LIS used in this study. Hence, in the future, a more involved study involving both space- and ground-based lightning observations over a longer period of time is required to provide a clearer picture. In addition, a future study can also attempt to demonstrate the impact of global warming on the multi-decadal weakening of lightning using high resolution numerical model-based experiments on an event scale during two different years spaced by at least 10-15 years so as to give a more prominent insight into the proposed hypothesis.

## Author Contribution

RC performed the complete analysis. PSM helped with the analysis and wrote the entire manuscript. RC helped in editing and supervision.

## Declaration of competing interests

The authors declare that they have no known competing financial interests or personal relationships that could have appeared to influence the work reported in this paper.

## Acknowledgements

The authors are grateful to NASA and ECMWF for making the TRMM lightning passes, meteorological reanalysis and modelled aerosol products available to the research community. Rohit Chakraborty is grateful to the Ministry of Science and Technology, Government of India, for providing support in the form of a CV Raman Post-Doc Fellowship and INSPIRE faculty research grant (registration no. DST/INSPIRE/04/2019/002096 and INSPIRE faculty code IFA19-EAS79). The authors thank Marikundam Harshitha from Arizona State University and Sodunke Mobolaji Aduramo of Moshood Abiola Polytechnic for their valuable support and suggestions.

## Data Availability

ERA5 reanalysis data is available at Copernicus Climate Change Service (C3S) Climate Data Store of ECMWF (https://cds.climate.copernicus.eu/cdsapp#!/home). High-resolution lightning datasets for the present study have been obtained from LIS archives of the NASA Global Hydrology Resource Center DAAC, USA (https://ghrc.nsstc.nasa.gov/lightning/data/data_lis_trmm.html  GHRC DAAC, 2020). The Aerosol components were utilized from the Modern-Era Retrospective analysis for Research and Applications version 2 provided by NASA (https://gmao.gsfc.nasa.gov/reanalysis/MERRA-2/ ). Finally, the future projections of lightning properties are taken from 5 general circulation model (GCMs) in Coupled Model Intercomparison Project (CMIP5) archive (https://esgf-node.llnl.gov/search/cmip5/ from Department of Energy, Lawrence Livermore National Laboratory, 2021).